\documentclass[twocolumn, showpacs,preprintnumbers,amsmath,amssymb]{revtex4}

\usepackage[pdftex]{graphicx}
\usepackage{subfigure}

\begin{document}

\addtolength{\textheight}{1.2cm}
\addtolength{\topmargin}{-0.5cm}

\newcommand{\etal} {{\it et al.}}

\title{Detection of entanglement between collective spins}

\author{F. Troiani}
\affiliation{S3 Istituto Nanoscienze, Consiglio Nazionale delle Ricerche, I-41100 Modena, Italy}

\author{S. Carretta}
\affiliation{Dipartimento di Fisica e Scienze della Terra, Universit\'a di Parma, I-43124 Parma, Italy}

\author{P. Santini}
\affiliation{Dipartimento di Fisica e Scienze della Terra, Universit\'a di Parma, I-43124 Parma, Italy}

\date{\today}

\begin{abstract}

Entanglement between individual spins can be detected by using thermodynamics quantities as entanglement witnesses. This applies to collective spins also, provided that their internal degrees of freedom are frozen, as in the limit of weakly-coupled nanomagnets. 
Here, we extend such approach to the detection of entanglement between subsystems of a spin cluster, beyond such weak-coupling limit. The resulting inequalities are violated in spin clusters with different geometries, thus allowing the detection of zero- and finite-temperature entanglement. Under relevant and experimentally verifiable conditions, all the required expectation values can be traced back to correlation functions of individual spins, that are now made selectively available by four-dimensional inelastic neutron scattering.

\end{abstract}


\maketitle

A great effort has been devoted in the last years to the generation and detection of quantum entanglement in diverse physical systems \cite{Amico08}. One of the most practical means in the latter perspective is represented by entanglement witnesses \cite{Guhne09,Horodecki09}. These are observables that can be experimentally accessed in the system of interest, and whose expectation values can exceed given thresholds only in the presence of specific forms of entanglement. The violation of the corresponding inequalities allows the detection of entanglement, without requiring the derivation of the systems state, and with a variable amount of knowledge of the system Hamiltonian.
In spin systems, routinely measured thermodynamic quantities, such as magnetic susceptibility, can be used as entanglement witnesses \cite{Ghosh03,Brukner06,Wiesniak05}. In clusters of antiferromagnetically coupled spins, internal energy allows one to demonstrate the non full-separability of the equilibrium state at low temperatures \cite{Brukner04,Dowling04,Toth05,Siloi12}. Exchange energy also allows the detection of multipartite entanglement in qubit systems \cite{Guhne05,Guhne06} and in clusters formed by $s>1/2$ spins \cite{Troiani12}. In the same spirit, spin-squeezing inequalities can be used to demonstrate entanglement in the vicinity of relevant quantum states by means of collective observables only \cite{Sorensen01,Vitagliano11}. 

In all these cases, the relevant subsystems are represented by individual spins, such as those carried by single ions within a molecular nanomagnet \cite{Gatteschi}. Here we extend such approach to the detection of entanglement between subsystems (hereafter labeled $A$ and $B$) of a spin cluster. Each subsystem is formed by a finite number of individual spins ($s_i^\chi$) and can be described in terms of a collective spins ($S_\chi$, with $\chi =A,B$). In many cases of interest, these partial spin sums undergo non-negligible fluctuations ($\Delta S_\chi$) in the low-energy eigenstates of the spin cluster. Typical
examples are represented by dimers of weakly coupled molecular nanomagnets \cite{Wernsdorfer02,Hill03}, where entanglement between the molecules has so far been discussed only in terms of a simplified two-macrospin model \cite{Candini10,Troiani10}. The approach developed hereafter provides the possibility of extending such investigations to the case where the internal degrees of freedom of each subsystem cannot be disregarded ($\Delta S_\chi \neq 0$). 

It is also shown that, if the fluctuations of the partial spin sums $S_A$ or $S_B$ are small, all the relevant quantities can be expressed in terms of correlation functions between pairs of individual spins. This result potentially presents a practical relevance, for such correlation functions can be experimentally derived by four-dimensional inelastic neutron scattering \cite{Baker12}. The validity of the underlying physical assumption can be deduced from a qualitative knowledge of the system, and, most importantly, from the experimentally accessible spin-pair correlation functions. 

Molecular nanomagnets represent a varied class of spin clusters, whose physical properties can be widely tuned by chemical synthesis \cite{Gatteschi}. In particular, low-spin nanomagnets with dominant antiferromagnetic interaction can be regarded as prototypical examples of highly correlated, zero-dimensional quantum systems. These molecular spin clusters thus represent a suitable arena for the investigation of quantum entanglement. The understanding and control of different forms of entanglement (e.g., that between single ions of a single molecule, or that between molecular spins within a dimer) also represents a preliminary requirement for the use of molecular nanomagnets as spin cluster qubits \cite{Troiani11}.

The paper is organized as follows. In Sec. \ref{SecEnt} we derive the inequalities that allow to detect entanglement between the composite spins. In Sec. \ref{SecPro} such inequalities are applied to the investigation of entanglement within different bipartitions of prototypical spin clusters. In Sec. \ref{SecCon} we discuss the measurement of the relevant observables by means of inelastic neutron scattering and draw our conclusions.

\section{Entanglement between \protect\\ composite spins}\label{SecEnt}

In the present Section, we derive inequalities that allow to detect entanglement between two subsystems $A$ and $B$ of a spin cluster, formed by $N_A$ and $N_B$ individual spins, respectively. The approach for passing from individual to composite spins is outlined in Subsection \ref{subsec1}. Different entanglement inequalities are typically used to detect entanglement in low- and in high-spin systems. These two cases are thus treated separately (Subsections \ref{subsec2} and \ref{subsec3}, respectively). A possible generalization of the approach to the case of more than two subsystems is sketched in Subsection \ref{subsec4}.

\subsection{From individual to composite spins}\label{subsec1}

Our starting point is represented by a generic inequality which applies to the separable states of a two-spin system:
\begin{equation}\label{ineq01}
\langle W \rangle \ge \gamma (S_A,S_B) .
\end{equation}
Here, $W$ is the entanglement witness and the lower bound $ \gamma $ depends on the spin values. If $S_A$ and $S_B$ are composite, rather than individual spins, 
with $ {\bf S}_\chi = \sum_{i=1}^{N_\chi} {\bf s}_i^\chi $ (and $\chi = A,B$),
then their lengths are no longer an intrinsic property of the system, but rather 
state-dependent quantities. The above inequality can thus be used to detect entanglement between the subsystems $A$ and $B$ only if the system state $\rho_{AB}$ is defined within a subspace with given values of the partial spin sums $S_A$ and $S_B$. 
However, if the witness commutes with the partial spin sums,
\begin{equation} \label{commut}
[ W , {\bf S}_A^2 ] = [ W , {\bf S}_B^2 ] = 0 ,
\end{equation}
then Eq. \ref{ineq01} can be generalized as follows to an arbitrary state $\rho_{AB}$:
\begin{equation} \label{eq00}
\langle W \rangle \ge \sum_{k} p(S^A_k,S^B_k) \gamma (S^A_k,S^B_k) \equiv \overline{\gamma},
\end{equation}
where $p(S^A_k,S^B_k)$ is the probability corresponding to eigenvalues 
$S^A_k(S^A_k+1)$ and $S^B_k(S^B_k+1)$ of ${\bf S}_A^2$ and ${\bf S}_B^2$, 
respectively. 
Equation \ref{eq00}, which essentially results from the vanishing contribution to the expectation value $ \langle W \rangle $ of the coherences between subspaces corresponding to different values of $ (S_A,S_B) $, is proven as follows. In view of the above commutation relations (Eq. \ref{commut}), one can define a complete basis formed by the states $ | \alpha , w , S^A_k , S^B_k \rangle $. These are the simultaneous eigenstates of $ {\bf S}^2_A $, $ {\bf S}^2_B $, and $W$, with the index $\alpha$ denoting additional quantum numbers, if any, required to specify the state. 
The expectation value of the witness can thus be written as:
\begin{eqnarray}\label{eq77}
\langle W \rangle \!\!\! & = & \!\!\! 
\sum_\alpha \sum_{k} \sum_w  \langle \alpha , w , S^A_k , S^B_k | \rho_{AB} W | \alpha , w , S^A_k , S^B_k \rangle
\nonumber\\       \!\!\! & = & \!\!\! 
\sum_\alpha \sum_{k} \sum_w   \langle \alpha , w , S^A_k , S^B_k | \rho_{AB}  | \alpha , w , S^A_k , S^B_k \rangle w
\nonumber\\       \!\!\! & = & \!\!\! 
\sum_{k}  p ( S^A_k , S^B_k ) \langle W \rangle_k  \ge 
\overline{\gamma} ,
\end {eqnarray}
Here, $ \langle W \rangle_k \equiv {\rm tr} (\rho_{AB}^k W)$ is the expectation value of the entanglement witness performed within the subspace $k$, 
and $\rho_{AB}^k$ is the density matrix projected onto the subspace $k$ and normalized:
\begin{equation}
\rho_{AB}^k \equiv \frac{ \Pi_k \rho_{AB} \Pi_k }{ {\rm tr} (\rho_{AB} \Pi_k )} 
= 
\frac{ \Pi_k \rho_{AB} \Pi_k }{p ( S^A_k , S^B_k )} ,
\end{equation}
with $ \Pi_k \equiv \sum_\alpha \sum_w | \alpha , w , S^A_k , S^B_k \rangle\langle \alpha , w , S^A_k , S^B_k | $.
In the last line of Eq. \ref{eq77}, we made use of the intuitive fact that the separability of $\rho_{AB}$ implies the separability of each $\rho_{AB}^k$. 

Equation \ref{eq00} can always be used to demonstrate theoretically the presence of entanglement between the subsystems $A$ and $B$, for all the relevant quantities can be derived from the density matrix $\rho_{AB}$. 
From an experimental viewpoint, the use of Eq. \ref{eq00} for entanglement detection is limited to the cases where the lower bound $ \overline{\gamma} $ can be derived from expectation values of accessible observables. In particular, we consider here the case where these correspond to two-spin operators, such as 
$ {\bf s}_i^\chi \cdot {\bf s}_j^{\chi'} $ ($\chi, \chi' = A,B$).
In general, the projectors $\Pi_k$, whose expectation values appear in the expression of $ \overline { \gamma } $, are of higher order in the single-spin operators. However,
as shown in the following Subsection, $ \overline{\gamma}$ can be written as a function of $ \langle {\bf s}_i^\chi \cdot {\bf s}_j^{\chi'} \rangle $ ($\chi , \chi' = A , B$) in a number of cases of interest, and specifically in the limit of a weak coupling between the subsystems $A$ and $B$.

\subsection{Entanglement in low-S states}\label{subsec2}

For any separable density matrix $\rho_{AB}$ of two spins $S_A$ and $S_B$, the following inequality applies \cite{Brukner04,Toth05,notaA4}:
\begin{equation} \label{eqA01}
\langle W \rangle = \langle {\bf S}_A \cdot {\bf S}_B \rangle \ge - S_A S_B = \gamma (S_A , S_B).
\end{equation}
Here, the set of additional quantum
numbers that define the basis states (see Eq. \ref{eq77}) is given
by: $ \alpha = ( \tilde{S}^A_2 , \dots , \tilde{S}^A_{N_A-1} ,
\tilde{S}^B_2 , \dots , \tilde{S}^B_{N_B-1},M) $, with
$\tilde{\bf S}^{\chi =A,B}_l = \sum_{i=1}^l {\bf s}^{\chi}_i$
and $M$ the total spin projection along $z$.
The violation of this inequality implies entanglement between $S_A$ and $S_B$, and can be typically observed in pairs of antiferromagnetically coupled spins, whose equilibrium density matrix at low temperatures ($T \lesssim J_{AB}$) is close enough to a low-$S$ state ($S=0$, if $S_A=S_B$).

In the case where $S_A$ and $S_B$ are composite, rather than individual spins, Eq. \ref{eqA01} can be generalized to
\begin{equation} \label{eqA02}
\langle {\bf S}_A \cdot {\bf S}_B \rangle \ge - \sum_k \langle \Pi_k \rangle S^A_k S^B_k 
= - \overline {S_A S_B}
\end{equation}
along the lines defined in the previous Subsection, being 
$ [ {\bf S}_A \cdot {\bf S}_B  , {\bf S}_A^2 ] = 
  [ {\bf S}_A \cdot {\bf S}_B  , {\bf S}_B^2 ] = 0$ (Eq. \ref{commut}). 
Here, the terms on the left-hand side of the inequality can be expressed as combinations of the quantities $ \langle {\bf s}_i^A \cdot {\bf s}_j^B \rangle $, while $ \overline{S_A S_B} $ depends on the expectation values $ \langle \Pi_k \rangle = p(S^A_k , S^B_k) $ of the projector $ \Pi_k $. 

\subsubsection{Approximate solutions}

In order to simplify the expression of the above lower bound, and to express it as a function of two-spin expectation values, we consider the limit of weak coupling between the subsystems $A$ and $B$. In such limit, the interaction with $B$ ($A$) induces a finite, but limited amount of $S$ mixing in the subsystem $A$ ($B$), and approximate expressions can be derived in two steps. 

As a first step, the average $ \overline{S_A S_B} $ is majorized by a function of $ \overline{S_\chi} $ and $ \overline{S^2_\chi} $ ($\chi = A,B$). This is performed by exploiting the Cauchy-Schwarz inequality \cite{Bhattacharya}:
\begin{equation}\label{eq06}
\overline{S_A S_B}
\le \overline{S_A} \, \overline{S_B} +
\left\{ \left( \overline{S_A^2} - \overline{S_A}^2 \right)
        \left( \overline{S_B^2} - \overline{S_B}^2 \right) \right\}^{1/2} ,
\end{equation}
which applies to any joint probability distribution of two real variables, such as $ p ( S^A_k , S^B_k ) $.
If the two subsystems $A$ and $B$ are identical, then
$\overline{S_A}=\overline{S_B}$
and
$\overline{S_A^2}=\overline{S_B^2}$.
The combination of Eq. \ref{eqA01} and Eq. \ref{eq06} thus gives:
\begin{eqnarray} \label{eq07}
\langle {\bf S}_A \cdot {\bf S}_B \rangle
\ge - \overline{S_\chi^2},
\end{eqnarray}
with $ \chi = A,B $.
We note that, when the ground state of the spin Hamiltonian is a singlet, the above inequality detects entanglement, if any. In fact, for $S=0$ the values of $S_A$ and $S_B$ have to be perfectly correlated \cite{notaA3}: this implies that 
$ \overline{S_A^2} = \overline{S_B^2} = \overline{S_A S_B} $.
The vanishing of the total spin $S$ also implies
$ \langle {\bf S}_A \cdot {\bf S}_A \rangle =
  \langle {\bf S}_B \cdot {\bf S}_B \rangle =
- \langle {\bf S}_A \cdot {\bf S}_B \rangle $.
Therefore, the inequality Eq. \ref{eq07} reduces to:
\begin{equation} \label{eq18}
\langle {\bf S}_A \cdot {\bf S}_B \rangle = -
\langle {\bf S}_\chi \cdot {\bf S}_\chi \rangle =
- \overline{S_\chi^2} - \overline{S_\chi} \ge - \overline{S_\chi^2} ,
\end{equation}
with $\chi = A,B$.
If the singlet ground state results from the product of two singlet states ($S_A=S_B=0$), the inequality is fulfilled. In all the other cases, the subsystems $A$ and $B$ are entangled, and such entanglement is detected by the violation of Eq. \ref{eq09}, being 
$ \overline{S_A} = \overline{S_B} > 0 $.

In the second step of our procedure, the averages $\overline{S_\chi^2}$ are replaced by simple functions of the expectation values $ \langle {\bf S}_\chi^2 \rangle $, and thus of $ \langle {\bf s}_i^\chi \cdot {\bf s}_j^\chi \rangle $.
The specific function to be used depends on whether the values of $ S_\chi $ approach their theoretical minimum ($0$ or $1/2$) or maximum ($\sum_{i=1}^{N_\chi} s_i^\chi$). These two cases are considered separately hereafter.

\par\bigskip

\paragraph{Low-spin subsystems.}
If the partial spin sum takes integer values, its smallest possible value is $S_\chi = 0$,
with $\chi = A,B$. We consider the case where the only relevant terms in the probability distribution
$ p(S^A_k,S^B_k) $ are $ p(0,0) \equiv \alpha $, $ p(0,1) = p(1,0) \equiv \beta $, and $ p(1,1) = 1 - \alpha - 2\beta $.
The corresponding expressions of the averages are
$ \overline{S_\chi} = \overline{S_\chi^2} = 1-\alpha-\beta $,
while
$ \langle {\bf S}_\chi^2 \rangle = 2(1-\alpha-\beta) $.
Therefore,
$ \overline{S_\chi^2} = \langle {\bf S}_\chi^2 \rangle / 2 $ and Eq. \ref{eq07} becomes:
\begin{eqnarray} \label{eq09}
\langle {\bf S}_A \cdot {\bf S}_B \rangle
\ge - \langle {\bf S}_\chi^2 \rangle / 2.
\end{eqnarray}

If the partial spin sum has half-integer values, then its lowest value is $S_\chi = 1/2$. One can proceed as in the previous case, and define:
$ p(1/2,1/2) \equiv \alpha $,
$ p(1/2,3/2) = p(3/2,1/2) \equiv \beta $,
and
$ p(3/2,3/2) = 1 - \alpha - 2\beta $.
The averages thus read
$ \overline{S_\chi} = 3/2 - (\alpha +\beta) $
and
$ \overline{S_\chi^2} = 9/4 - 2(\alpha +\beta) $,
while
$ \langle {\bf S}_\chi^2 \rangle = 15/4-3( \alpha + \beta ) $.
This results in the relation
$ \overline{S_\chi^2} = 2 \langle {\bf S}_\chi^2 \rangle / 3 - 1/4 $.
Equation \ref{eq07} can thus be expressed in the form:
\begin{eqnarray} \label{eq10}
\langle {\bf S}_A \cdot {\bf S}_B \rangle
\ge - 2 \langle {\bf S}_\chi^2 \rangle / 3 + 1/4.
\end{eqnarray}

\par\bigskip

\paragraph{High-spin subsystems.}
If the partial spin sum tends to take the maximum possible value, $ S_M = \sum_{i=1}^{N_\chi} s_i^\chi $, then the limit of small $S$-mixing for the subsystems corresponds to the probability distribution
$ p(S_M,S_M) \equiv \alpha $, $ p(S_M,S_M-1) = p(S_M-1,S_M) \equiv \beta $,
and
$ p(S_M-1,S_M-1) = 1 - \alpha - 2\beta $.
In this case, the averages and expectation values are
$ \overline{S_\chi} = S_M - 1 + \alpha + \beta $
and
$ \overline{S_\chi^2} = (S_M-1)^2 + (2S_M-1)(\alpha + \beta) $,
while
$ \langle {\bf S}_\chi^2 \rangle = S_M ( S_M - 1 ) + 2S_M ( \alpha + \beta ) $.
Equation \ref{eq07} thus takes the form:
\begin{eqnarray} \label{eq11}
\langle {\bf S}_A \cdot {\bf S}_B \rangle
& \ge & - (S_M-1)^2 \nonumber\\
& & - \frac{S_M-1/2}{S_M} [\langle {\bf S}_\chi^2 \rangle - S_M(S_M-1)].
\end{eqnarray}

\subsection{Entanglement in high-S states}\label{subsec3}

According to the spin-squeezing inequalities for arbitrary spins \cite{Vitagliano11}, the separability of the density matrix $\rho_{AB}$ of $S_A$ and $S_B$ implies:
\begin{equation} \label{eq21}
2 \langle {\bf S}_A \cdot {\bf S}_B \rangle -
    4\langle S_z^A S_z^B \rangle
\le S_A^2+S_B^2 - \langle S_z \rangle^2 .
\end{equation}
This inequality is violated by those entangled states with a high (i.e. close to $S_A+S_B$) value of the total spin $S$ and a small modulus of its projection along $z$. The simplest example of such a state might be represented by the spin triplet, with vanishing projection along $z$, formed by two $s=1/2$ spins.

Proceeding as in the previous Subsection, we generalize Eq. \ref{eq21} to the case where $S_A$ and $S_B$ are not individual spins, but rather partial spin sums, corresponding to generic subsystems $A$ and $B$:
\begin{equation} \label{eq22}
2 \langle {\bf S}_A \cdot {\bf S}_B \rangle -
4 \langle S^A_z S^B_z \rangle  
\le 
\sum_k \langle \Pi_k \rangle
[ (S_k^A)^2 + (S_k^B)^2 - \langle S_z \rangle^2_k ] ,
\end{equation}
where $\Pi_k$ is the projector on the subspace $ ( S^A_k , S^B_k ) $.

\subsubsection{Approximate solutions}

We first consider the limit where the occupation of all subspaces but the ones corresponding to 
$ S = S_M^A + S_M^B = \sum_{i=1}^{N_A} s^A_i + \sum_{i=1}^{N_B} s^B_i $ 
can be neglected. If $A$ and $B$ are formed by the same number of identical spins $s_i=s$, then $S^A_M=S^B_M=Ns/2 \equiv S_M$. The inequality Eq. \ref{eq22} thus becomes:
\begin{equation}
\langle S_z^A S_z^B \rangle \ge \langle S_z \rangle^2 / 4 .
\end{equation}
Any state violating such inequality is entangled. If the system Hamiltonian is invariant with respect to time-reversal symmetry, then $ \langle S_z \rangle_k = 0 $ for any thermal state. The inequality is thus violated in all the cases where the $z$ components of ${\bf S}_A$ and ${\bf S}_B$ are anticorrelated.

If the state of each subsystem includes small contributions from the subspaces $S_\chi=S_M^\chi -1$ and the subsystems $A$ and $B$ are identical, then the approximate expressions of $\overline{S_\chi^2}$ lead to the inequality:
\begin{eqnarray}\label{eqappr}
2\langle {\bf S}_A \cdot {\bf S}_B \rangle\!\!\! &-&\!\!\! 4\langle S_z^A S_z^B \rangle
\le 2 (S_M-1)^2 \nonumber\\ & &
+ \frac{2S_M-1}{S_M}  [\langle {\bf S}_\chi^2 \rangle - S_M(S_M-1)].
\end{eqnarray}

\subsection{Entanglement between more than two composite spins}\label{subsec4}

The proocedure outlined in the previous Subsections can be generalized to the case of $n$ composite spins $S_{A_i}$. The starting point is represented by a generic inequality which applies to the fully separable states of an $n$-spin system:
\begin{equation}\label{eqB01}
\langle W \rangle \ge \gamma (S_{A_1} , S_{A_2}, \dots , S_{A_n}) ,
\end{equation}
where the lower bound $ \gamma $ depends on the spin values. In the case where the $S_{A_i}$ are composite, rather than individual spins, 
with $ {\bf S}_\chi = \sum_{i=1}^{N_\chi} {\bf s}_i^\chi $ (and $\chi = A_1, \dots , A_n$),
the above inequality allows the detection of entanglement only if the system state is defined within a subspace $k$ with given values of the partial spin sums $S_{A_i}$. 
However, if 
\begin{equation}
[ W , {\bf S}_{A_i}^2 ] = 0 \ (i=1, \dots , n),
\end{equation}
then Eq. \ref{eqB01} can be generalized as follows:
\begin{equation} \label{eq222}
\langle W \rangle \ge \sum_{k} p(S_k^{A_1}, \dots , S_k^{A_n}) \gamma (S_k^{A_1}, \dots , S_k^{A_n}) \equiv \overline{\gamma},
\end{equation}
where $ p(S_k^{A_1}, \dots , S_k^{A_n}) = \langle \Pi_k \rangle $ is the probability corresponding to a subspace $k$, with defined values of the partial spin sums.

As a representative example, we consider the case of a spin cluster with a singlet ground state. At temperatures lower than the typical exchange constant, the density matrix of such a spin cluster typically violates the following inequality \cite{Wiesniak05}:
$
\langle {\bf S} \cdot {\bf S} \rangle \ge \sum_{j=1}^{n} S_j ,
$
which is instead fulfilled by any fully separable density matrix. If the individual spins are replaced by composite ones ${\bf S}_{A_i}$, entanglement between the subsystems $A_i$ can be detected by observing the violation of the inequality: 
\begin{equation} \label{eq33}
\langle {\bf S} \cdot {\bf S} \rangle \ge \sum_k \langle \Pi_k \rangle 
\sum_{i=1}^{n} {S^{A_i}_k} .
\end{equation}
Approximate solutions for expressing the averages $ \overline{S}_j$ in terms of expectation values can be derived in the small $S$-mixing limit, along the same lines as in the case of two subsystems.

\section{Bipartitions~of~prototypical spin~clusters}
\label{SecPro}

\begin{figure}[ptb]
\begin{center}
\includegraphics[width=8.5cm]{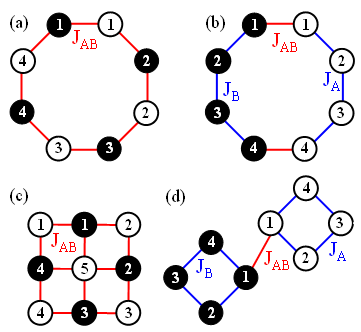}
\end{center}
\caption{(color online) Geometries and bipartitions of the considered spin clusters. (a,b) The system is an $N$-spin ring, with antiferromagnetic exchange interaction between nearest neighbors. The two equivalent subsystems $A$ (white circles) and $B$ (black) are formed by $N/2$ consecutive (a) or alternate spins (b) each. (c) The system consists of a spin grid, with two inequivalent sublattices $A$ and $B$. (d) Dimer-like structure, where the subsystems $A$ and $B$ are given by exchange-coupled spin rings.}
\label{FigStruct}
\end{figure}

The inequalities discussed so far are suited for systems that can be partinioned in two complementary subsystems ($A$ and $B$), such that the partial spin sums $S_A$ and $S_B$ are characterized by finite but small fluctuations in the system ground state. This might occur in clusters of exchange-coupled spins, if the typical exchange constant between the spins of each subsystem ($J_A$ and $J_B$) is much larger than that between spins belonging to different subsystems ($J_{AB}$). Molecular nanomagnets provide systems of this kind, such as supramolecular assemblies, formed by weakly coupled molecules. In Subsection \ref{subsecdimer} we discuss a prototypical sistem of this kind, represented by a ring dimer. 
The weak-coupling regime, defined by the inequality $ |J_{AB}| \ll |J_A|,|J_B|$, doesn't however represent a necessary condition for $S_A$ and $S_B$ to be approximately good quantum numbers. This feature also shows up in spin clusters with a ring or a grid geometry. The first of such systems is discussed in Subsection \ref{subsecrings}, where, depending on the partition of the ring, the intermediate ($ |J_{AB}| = |J_A|,|J_B|$) or strong-coupling ($ J_A = J_B = 0 $) regimes apply. The latter regime also characterizes the spin grid with nearest-neighbor interactions, which is discussed in Subsection \ref{subsectiongrid}.

\subsection{Entanglement in dimer-like spin clusters}\label{subsecdimer}

Molecular nanomagnets can act not only as individual quantum systems, but also as weakly coupled units within supramolecular structures \cite{Wernsdorfer02,Hill03,Candini10}. Hereafter, we consider a prototypical dimer-like structure (Fig. \ref{FigStruct}(d)), formed by two spin rings (labelled $A$ and $B$). The spin Hamiltonian of the dimer, 
$ H = H_A + H_B + H_{AB} $, includes two kinds of contributions, corresponding to the intra- and inter-ring exchange interactions:
\begin{equation}\label{eqdimer}
H_{\chi =A,B}= J_\chi \sum_{i=1}^{N_\chi} {\bf s}_i^\chi \cdot {\bf s}_{i+1}^\chi, \
H_{AB} = J_{AB} {\bf s}_1^A \cdot {\bf s}_1^B .
\end{equation}
In order for the ground state of $H$ to exhibit entanglement between the $A$ and $B$ rings, the inter-molecular exchange has to be antiferromagnetic ($J_{AB}>0$). If $J_A=J_B$ and $N_A=N_B$, the ground state of $H$ corresponds to a spin singlet.

\subsubsection{Antiferromagnetic intra-ring interactions}

\begin{figure}[ptb]
\begin{center}
\includegraphics[width=8.5cm]{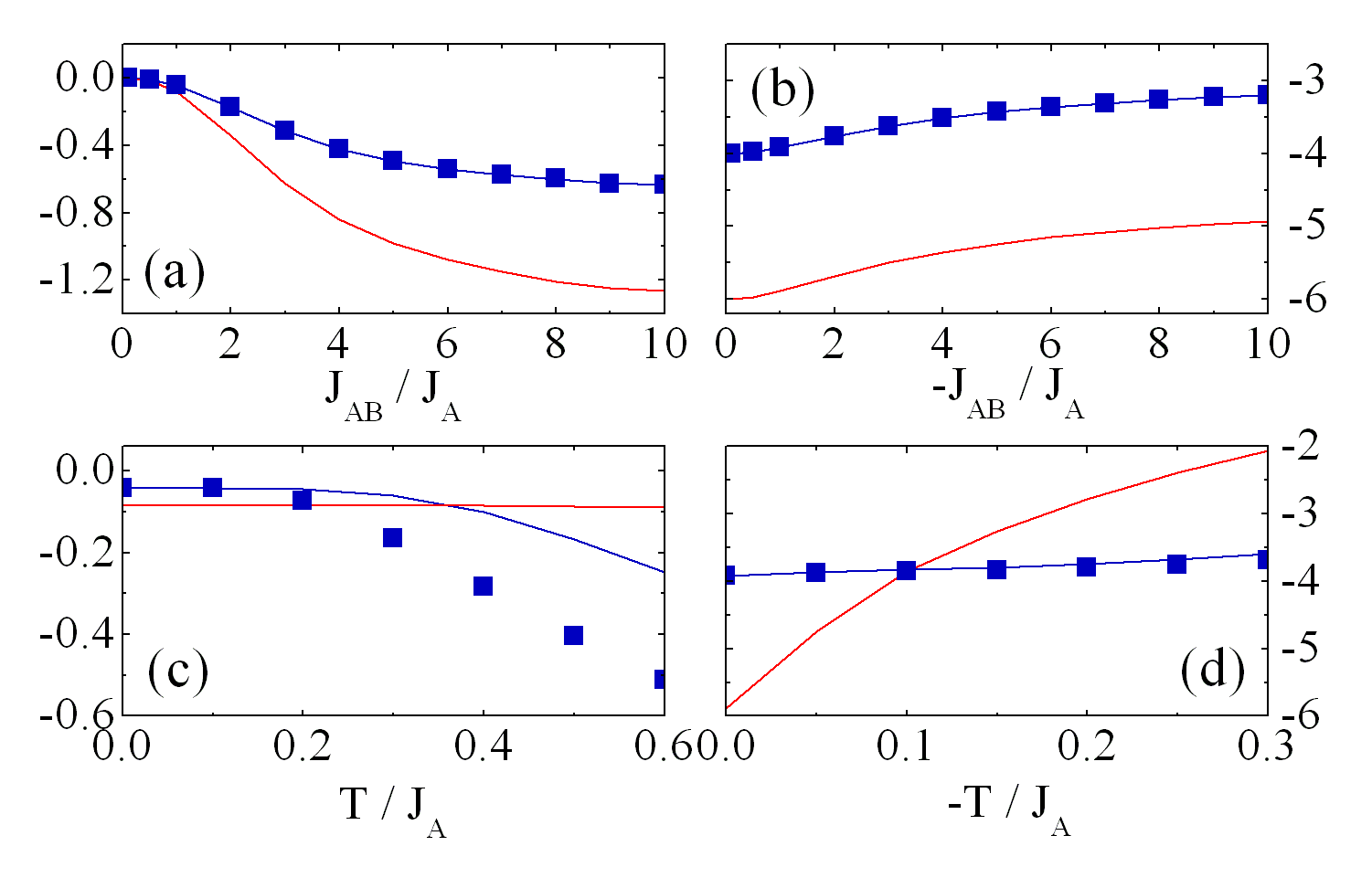}
\end{center}
\caption{(color online) (a) Dependence on the ratio $J_{AB}/J_A$ of $ \langle {\bf S}_A \cdot {\bf S}_B \rangle $ (red curves) and $ -\overline{S_AS_B} $ (blue curves) for the ground state of the two antiferromagnetically coupled rings (Eq. \ref{eqdimer}). The blue squares correspond to the approximate expression of $ -\overline{S_AS_B} $. The two rings are identical, with $J_A=J_B>0$. 
(b) Same as above, but with ferromagnetic intra-ring interactions ($J_A=J_B<0$).
(c) Temperature dependence of $ \langle {\bf S}_A \cdot {\bf S}_B \rangle $ and $ -\overline{S_AS_B} $ (exact and approximate expressions), for $J_A=J_B=J_{AB}>0$. 
(d) Same as above, but with ferromagnetic intra-ring interactions ($J_A=J_B=-J_{AB}<0$).}
\label{FigDimer}
\end{figure}

We start by considering the case where the intra-ring exchange is antiferromagnetic ($J_A=J_B>0$, with $N_A=N_B=4$), such that the ground state of $H_\chi$ corresponds to $S_\chi = 0$. In the limit $ J_{AB} \ll J_A $, the partial spin sums $S_A$ and $S_B$ are good quantum numbers, and the ground state tends to the product of the two singlet states $S_A=S_B=0$. For larger values of $J_{AB} / J_A$, the interaction between $s_1^A$ and $s_1^B$ induces an increasing occupation of states with $S_A,S_B >0$. In the limit $J_{AB} \gg J_A$, $s_1^A$ and $s_1^B$ form a singlet, while the remaining three spins of each ring form two uncoupled trimers. The comparison between $ \langle {\bf S}_A \cdot {\bf S}_B \rangle $ (Fig. \ref{FigDimer}(a), red curve) and $ -\overline{S_A S_B} $ (blue curve) shows that, the ground state exhibits inter-ring entanglement in both these limits, and in all the intermediate cases. Besides, the averages $ -\overline{S_A S_B} $ are very well approximated by the expression reported in Eq. \ref{eq09} (blue squares). The ground state of the dimer thus presents entanglement between the subsystems $A$ and $B$ for arbitrary values of $J_{AB}/J_A$, and such entanglement can always be detected by the inequality Eq. \ref{eqA02}, expressed in terms of spin-pair correlation functions. An analogous result (not shown) has been obtained for a ring dimer defined by the spin Hamiltonian $H$ (Eq. \ref{eqdimer}), formed by inequivalent spins: $s_1^\chi = 1$, $s_{i>1}^\chi = 3/2$ ($\chi =A,B$). This can be regarded as a simplified model of the (Cr$_7$Ni)$_2$ dimers, that have been recently synthesized in a number of different derivatives \cite{Candini10}.

In order to investigate the robustness of entanglement with respect to temperature, we compute the dependence on $T$ of $ \langle {\bf S}_A \cdot {\bf S}_B \rangle $ (panel (c), red curve) and $ -\overline{S_AS_B} $ (blue curve), for $J_{AB}=J_A$. The inequality Eq. \ref{eqA02} is violated ut to the threshold temperature $T=0.366\, J_A$, corresponding to about half of the gap $\Delta$ between the ground state singlet and first excited triplet of the dimer. The approximate expression reported in Eq. \ref{eq09} (blue squares) slightly underestimates $ -\overline{S_\chi^2}$ in the relevant temperature range, thus underestimating the threshold temperature. The negativity \cite{Guhne09} has a small value ($\mathcal{N}=0.159$) for $T=0.366\, J_A$, and vanishes for $T=1.05\, J_A$.

\subsubsection{Ferromagnetic intra-ring interactions}

An antiferromagetic interaction between the spins $s_1^A$ and $s_1^B$ (Eq. \ref{eqdimer}) also tends to entangle the two rings in the presence of ferromagnetic intra-ring interactions ($J_A=J_B<0$). In this case, the ground state in the limit $J_{AB} \ll |J_A|$ corresponds to a singlet, with $S_A=S_B=2$. Values of $J_{AB} / |J_A|$ up to 10 lead to a limited amount of $S-$mixing within each ring. As in the previous case, the inequality Eq. \ref{eqA02} is always violated, showing that the two rings are always entangled in the low-temperature limit (Fig. \ref{FigDimer}, panel (b)). Also, the approximate expression Eq. \ref{eq11} approaches the exact value of $ -\overline{S_A S_B} $, thus enabling the detection of entanglement between the rings in terms of the spin-pair correlation functions.

The inequality Eq. \ref{eqA02} allows the detection of entanglement up to temperatures of about $0.1\, |J_A|$ ($J_{AB}=-J_A>0$, panel (d)). In the relevant temperature range, the average $-\overline{S_AS_B}$ is very well approximated by the expression in Eq. \ref{eq11} (blue squares). Both quantities however underestimate the temperature range where the rings are actually entangled, being the negativity finite up to $ T \simeq |J_A|$.

\subsection{Subsystems within a spin ring}\label{subsecrings}

The two-macrospin model also applies to spin clusters that don't display any dimer-like structure, and thus don't lend themselves to be naturally partitioned into two (weakly-coupled) subsystems. As a first example of this different kind of systems, we consider the case of single spin rings. Molecular nanomagnets provide a variety of ring-like structures, with different spin numbers ($N$) and values ($s$) \cite{Gatteschi}. Hereafter, we focus on a ring formed by identical spins, with antiferromagnetic interaction between nearest neighbors. We consider two possible partitions of such ring into two equivalent subsystems.

\subsubsection{Entanglement between even- and odd-numbered spins}

In the first partition we consider, the two subsystems are formed by the odd- and even-numbered spins, respectively, and $N=8$ [Fig. \ref{FigStruct}(a)]. The three terms of the spin Hamiltonian $H=H_A+H_B+H_{AB}$ thus read:
\begin{equation}
H_{AB} = J_{AB} \sum_{i=1}^4 ({\bf s}_i^A + {\bf s}_{i+1}^A ) \cdot {\bf s}_i^B , \ H_A=H_B=0 ,
\end{equation}
where ${\bf s}_5^A \equiv {\bf s}_1^A$
and
$
{\bf S}_\chi = \sum_{i=1}^4 {\bf s}_i^\chi
$. 
In the ground state of $H$, the spins ${\bf s}^A_i$ tend to be parallel to each other (and antiparallel to the ${\bf s}^B_j$), and each partial spin sum tends to the theoretical maximum $Ns/2$.

\begin{figure}[ptb]
\begin{center}
\includegraphics[width=\columnwidth]{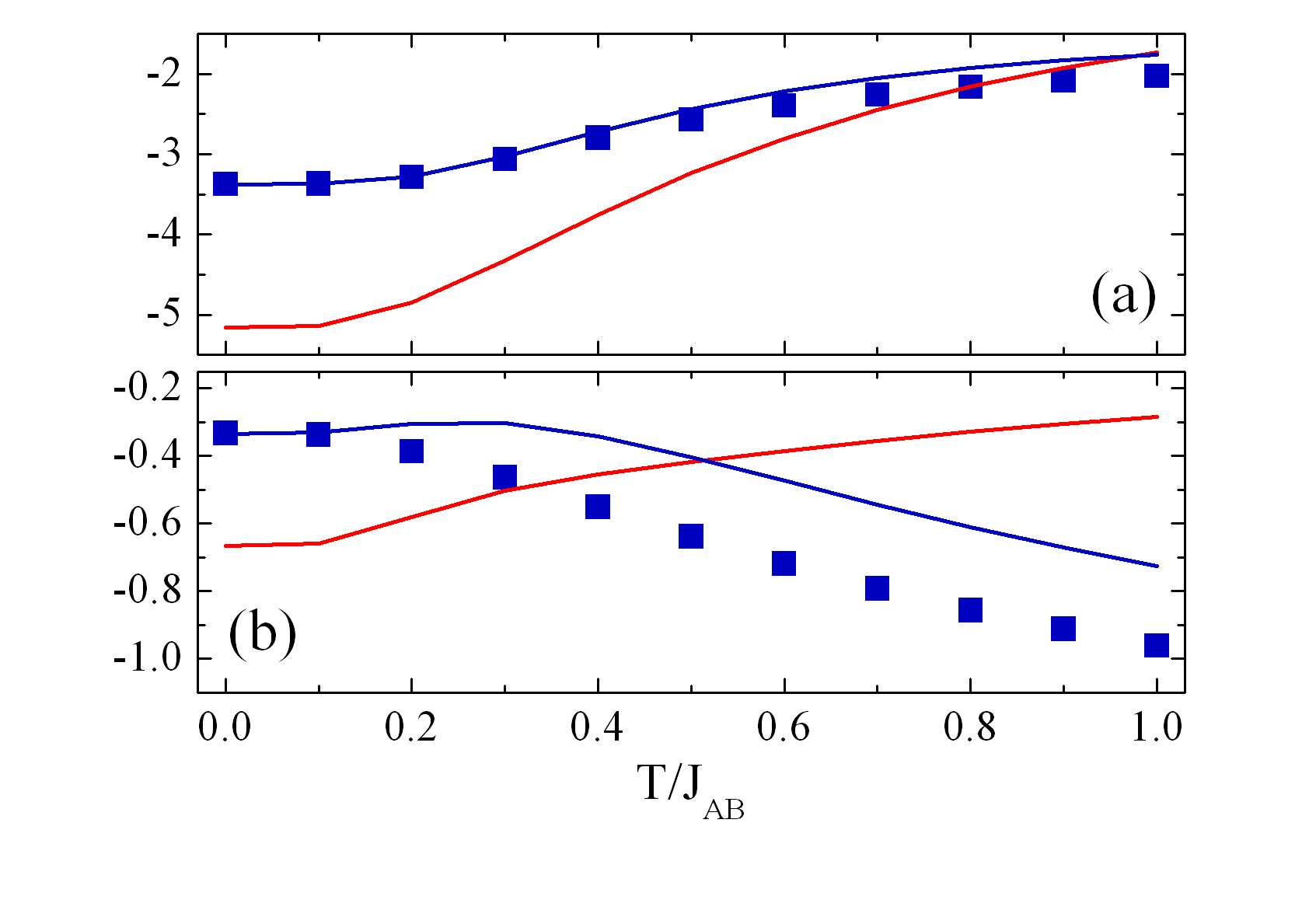}
\end{center}
\caption{(color online) Temperature dependence of
$\langle {\bf S}_A \cdot {\bf S}_B \rangle $ (red curves),
and
$ -\overline{S_A S_B} $ (blue) for the ring formed by $N=8$ rings $s=1/2$.
The inequality detects entanglement between the subsystems $A$ and $B$ in the temperature range where
$\langle {\bf S}_A \cdot {\bf S}_B \rangle < - \overline{S_A S_B} $.
The panels (a) and (b) refer to the partitions represented in Fig. \ref{FigStruct}, panels (a) 
and (b), respectively. The blue squares correspond to the approximate expressions of the thresholds in the large-$S_\chi$ (a) and small-$S_\chi$ limits (b).}
\label{FigRing}
\end{figure}

The calculations presented hereafter refer to the case of $N=8$ spins $s=1/2$. The average length of the intermediate spins in the system ground state is $ \overline{S_A} = \overline{S_B} = 1.79 $, corresponding to an entangled ground state \cite{notaA2,Waldmann01}. 
In order to verify the robustness of such entanglement with respect to temperature, we report the dependence on $T$ of $\langle {\bf S}_A \cdot {\bf S}_B \rangle$ and $ -\overline{S_AS_B} $ (Fig. \ref{FigRing}(a), red and blue curves, respectively). The inequality Eq. \ref{eqA02} is violated for $ T \lesssim J_{AB} $, which is approximately twice as large as the gap $ \Delta $ between the ground state and the first excited triplet. Besides, in the considered temperature range, the threshold value $ -\overline{S_AS_B} $  remains close to the approximate expression of spin-pair correlation functions given in Eq. \ref{eq11} (blue squares). The threshold temperature derived by means of Eq. \ref{eqA02} compares well with the temperature dependence of the negativity $\mathcal{N}$ \cite{Guhne09}. In fact, $\mathcal{N}=0.125$ at $T=J_{AB}$, whereas $\mathcal{N}=0$ for $T \gtrsim 1.35 \, J_{AB}$. We finally note that entanglement between sublattices in the spin ring persists to higher temperatures with respect to that between neighboring spins, which vanishes for $ T \gtrsim 0.8\, J_{AB}$. 

\subsubsection{Entanglement between consecutive spin segments}

In the second partition we consider, the subsystems $A$ and $B$ are formed by the first and second four spins, respectively [Fig. \ref{FigStruct}(b)]:
\begin{equation}
H_\chi=J_\chi \sum_{i=1}^3 {\bf s}_i^\chi \cdot {\bf s}_{i+1}^\chi ,\
H_{AB} = J_{AB} ( {\bf s}_1^A \cdot {\bf s}_1^B + {\bf s}_4^A \cdot {\bf s}_4^B ) ,
\end{equation}
where $ \chi = A,B$ and $J_A=J_B=J_{AB}$.
The ground state of $H_\chi$ is thus a spin singlet ($S_\chi=0$), while $H_{AB}$ mixes the $S_A=S_B=0$ state with those corresponding to finite values of the partial spin sums. In particular, for $s_i^\chi =1/2$ we obtain $\overline{S_\chi}=0.333$ \cite{notaA1}. This results in a violation of Eq. \ref{eqA02}, even though a less prominent one with respect to the one obtained with the previous partition of the ring. 

As to the effect of temperature, entanglement between these two subsystems is detected by the above criterion up to $T \lesssim 0.5 \, J_A \simeq \Delta $, where the value of $ \langle {\bf S}_\chi \cdot {\bf S}_\chi \rangle $ (Fig. \ref{FigRing}(b), red curve) becomes larger than the threshold $-\overline{S_A S_B}$ (blue). Unlike the case of the previous partition, the approximate expression of Eq. \ref{eq09} underestimates the threshold value in the relevant temperature range (the discrepancy arises mainly from the difference between $\overline{S_AS_B}$ and $\overline{S_A^2}=\overline{S_B^2}$, not shown). The use of such expression thus allows to detect entanglement in terms of expectation values of spin-pair operators, but only in the temperature range $T \lesssim 0.3\, J$. The above results suggest that entanglement between the two segments that form the ring is less robust with respect to temperature than that between even- and odd-numbered spins. This is confirmed by the temperature dependence of the negativity $\mathcal{N}$, that is finite ($\mathcal{N}=0.323$) at $T=0.5\, J$ and vanishes at $ T \simeq 0.975\, J$.

\subsection{Subsystems within a spin grid}\label{subsectiongrid}

Antiferromagnetic wheels don't represent the only nanomagnets whose ground state can be approximately described in terms of a two-macrospin model. Another class of systems that possess this property is represented by planar molecules such as the $ 3 \times 3 $ Cu \cite{Zhao00} and Mn \cite{Waldmann02} grids. 
In the following we focus on the former molecule, where each ion corresponds to an $s=1/2$ spin, and assume for simplicity a single exchange constant for all antiferromagnetic interactions between nearest neighbors. Within the spin grid, one can identify two sublattices, $A$ and $B$, such that each spin belonging to $A$ is coupled only to spins of $B$ [see Fig. \ref{FigStruct}(c)]. Unlike those of the spin ring, these subsystems are inequivalent, being $N_A \neq N_B$. The three terms of the spin Hamiltonian $H=H_A+H_B+H_{AB}$ read:
\begin{equation} \label{hamgrid}
H_{AB} = J_{AB} {\bf s}^A_5 \cdot {\bf S}_B +
\sum_{i=1}^4 {\bf s}^A_i \cdot ({\bf s}^B_{i-1} + {\bf s}^B_i ) , \ H_A\!=\!H_B\!=\!0 ,
\end{equation}
where ${\bf s}^B_0 \equiv {\bf s}^B_4 $. 
Within the $S=1/2$ ground doublet, each spin tends to be parallel to the ones of the same subsystem and antiparallel to those of the other one. In fact, the average values of the partial spin sums are given by $ \overline{S_A} = 2.31 $ and $ \overline{S_B} = 1.83 $. 

\begin{figure}[ptb]
\begin{center}
\includegraphics[width=8.5cm]{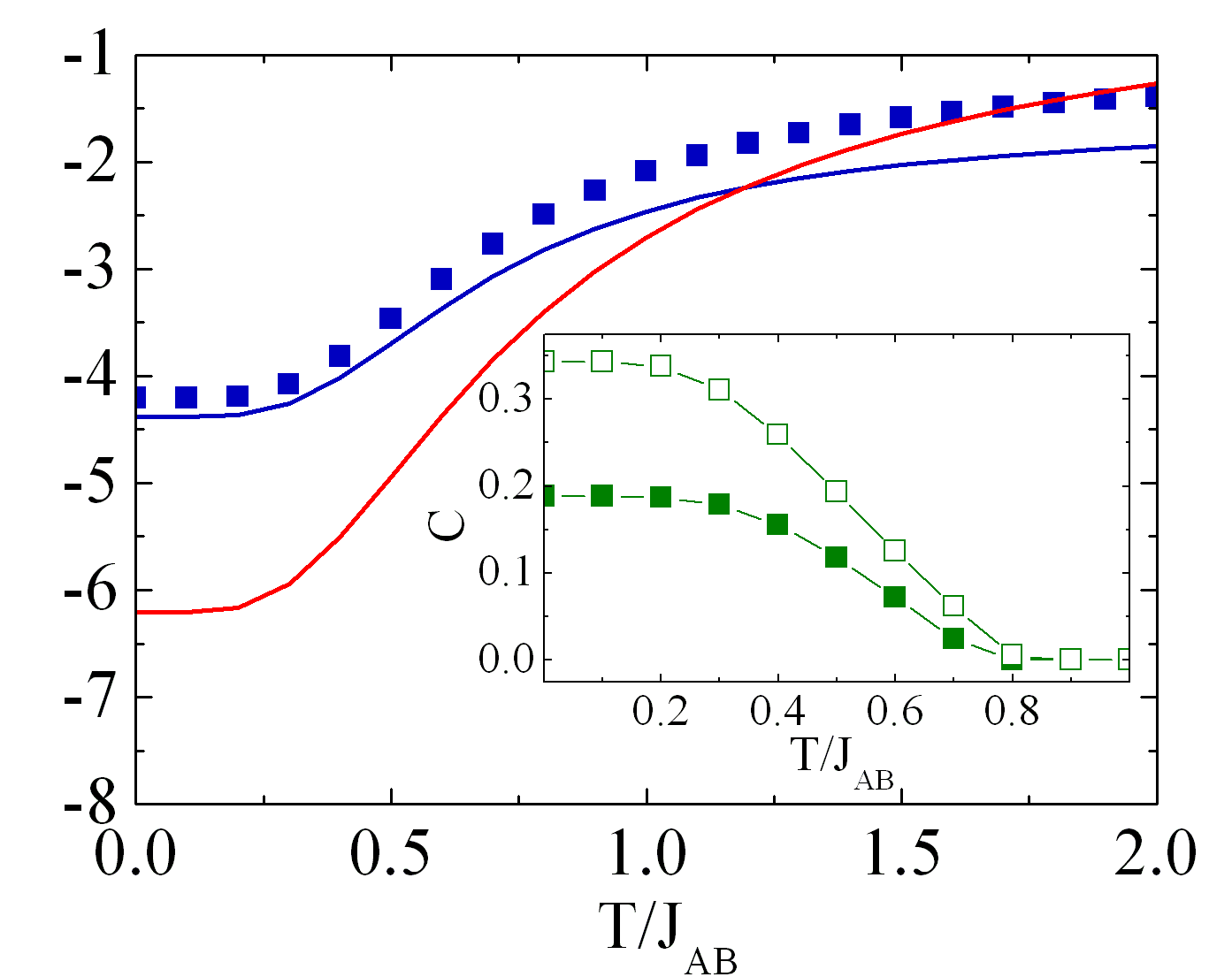}
\end{center}
\caption{(color online) Temperature dependence of
$\langle {\bf S}_A \cdot {\bf S}_B \rangle $ (red curve),
and
$ -\overline{S_A S_B} $ (blue) for the grid formed by $N=9$ spins $s=1/2$. The blue squares correspond to the approximate expressions of the thresholds in the large-$S_\chi$ 
limit.
Inset: Temperature dependence of the concurrence between the central spin $s^A_5$ and any of the $s^B_i$ (filled squares), and between $s^B_i$ and a neighboring $s^A_{j<5}$ (empty squares).}
\label{FigGrid}
\end{figure}
As in the previously considered spin clusters, the antiferromagnetic interactions between the spins $s_i^A$ and $s_j^B$ tend to entangle the two subsystems, and to induce the violation of the inequality Eq. \ref{eqA02}. In particular, the expectation value of 
$ \langle {\bf S}_A \cdot {\bf S}_B \rangle $ (Fig. \ref{FigGrid}, red curve) falls below the threshold $ -\overline{ S_A S_B } $ (blue) in the temperature range $ T \lesssim 1.38\, J_{AB}$. Along the lines of Sec. \ref{subsec2}, we approximate the threshold values with simple functions of the partial-spin sums. In the small $S_\chi-$mixing limit, one can disregard all the probabilities $p(S_k^A,S_k^B)$, but the following ones: 
$ p(5/2,2) = \alpha $, $ p(5/2,1) = \beta $, and $ p(3/2,2) = 1 - \alpha - \beta $. 
Expressing $\overline{S_AS_B}$, ${\bf S}_A^2$ and ${\bf S}_B^2$ in terms of these probabilities, and eliminating $\alpha$ and $\beta$, one obtains:
\begin{equation}
\langle {\bf S}_A \cdot {\bf S}_B \rangle 
\ge 
- 2 \langle {\bf S}_A^2 \rangle / 5 
- 5 \langle {\bf S}_B^2 \rangle / 8 
+ 9/4.
\end{equation}
The temperature dependence of the above lower bound is reported in Fig. \ref{FigGrid} (blue squares). The approximate expression slightly underestimates the value of the $\overline{S_AS_B}$. However, due to the small slope of the curves in the relevant region, this results in a larger relative error concerning the temperature range where entanglement persists.
In order to assess the suitability of inequality Eq. \ref{eqA02} to detect entanglement between the sublattices of the spin grid, we consider the temperature dependence of the negativity $\mathcal{N}$ \cite{Guhne09}. At the threshold temperature $T=1.38\, J_{AB}$, the negativity takes a finite, though small, value: $ \mathcal{N} = 0.0469$.  The negativity vanishes for slightly higher temperatures, and more specifically at $T \simeq 1.75\, J_{AB}$.

From a physical point of view, it might be instructive to compare the above temperature ranges with the ones that characterize spin-pair entanglement. In fact, the antiferromagnetic interactions in $H$ also tend to entangle neighboring spins within the grid. Such spin-pair entanglement is here quantified by the concurrence \cite{Hill97}, which is plotted in the figure inset as a function of temperature. The concurrence vanishes below $T \simeq J_{AB}$ for both the spin pairs along the sides of the grid (second contribution in Eq. \ref{hamgrid}, empty squares) and the pairs that involve the central spin $s_5^A$ (first contribution in Eq. \ref{hamgrid}, filled squares). The pairs of uncoupled spins are unentangled at all temperatures. Entanglement between the subsystems $A$ and $B$ is thus significantly more robust with respect to temperature than that between pairs of individual spins.

\section{Discussion and conclusions}\label{SecCon}

The inequalities derived above allow the detection of entanglement through equal-time spin-pair correlation functions. In fact, both the expectation values $ \langle {\bf S}_\chi \cdot {\bf S}_{\chi'} \rangle $ (with $ \chi , \chi' = A , B $) and (in a number of cases of interest) the averages $\overline{\gamma}$ can be expressed as linear combinations of the $ \langle {\bf s}_i^\chi \cdot {\bf s}_j^{\chi'} \rangle $. These expectation values can be experimentally accessed by magnetic neutron scattering, for the neutron cross-section can be written in terms of dynamical correlation functions between pairs of individual spins \cite{MarshallLovesey}. In particular, it was recently shown \cite{Baker12} that, if accurate data for the scattering function are available on a large-enough portion of the energy and momentum space, it is possible to extract equal-time two-spin correlation functions directly, without any prior knowledge of the system Hamiltonian. This allows the detection of entanglement between arbitrary subsystems also if the spin Hamiltonian cannot be reliably known, as is the case when the fit is not univocal, or even impractical, due to the size of the Hilbert space. It should also be noticed that neutron scattering requires samples containing a macroscopic number of identical spin clusters, and is thus well suited for the study of molecular nanomagnets.

Four-dimensional inelastic neutron scattering has recently been applied to the Cr$_8$ molecule \cite{Baker12}, essentially consisting in an octagon of $s=3/2$ spins. The equal-time correlation functions that have been measured provide a direct experimental demonstration of entanglement between the ring subsystems considered in Subection \ref{subsecrings}. 
In the low-temperature limit, the state of Cr$_8$ approximately coincides with a spin singlet 
($ \langle {\bf S} \cdot {\bf S} \rangle \simeq 0 $),  
whereas the partial spin sums 
$ \langle {\bf S}_A\cdot {\bf S}_A\rangle = 
  \langle {\bf S}_B\cdot {\bf S}_B\rangle \simeq 
- \langle {\bf S}_A\cdot {\bf S}_B\rangle $
approximtely correspond to 37 and 2 for the partitions represented in Fig. \ref{FigStruct} (a) and (b), respectively.
These values imply the violation of the
inequalities reported in Eqs. \ref{eq11} and \ref{eq09}. At low
temperatures, the subsystems $A$ and $B$ of Cr$_8$ are thus
entangled for both the above partitions.
Such an experimental investigation of entanglement can also be applied to dimers of weakly-coupled molecules, such as (Cr$_7$Ni)$_2$. These systems, that are of potential interest for quantum computation \cite{Troiani05} and simulation \cite{Santini11}, have in fact been synthesized in large high-quality crystals, that are well suited for neutron-scattering. In particular, the inequalities Eq. \ref{eqA02},\ref{eq09} can be used to prove the existence of entanglement between the two Cr$_7$Ni molecules, also if the inter-ring exchange $ J_{AB}$ is large enough to induce a small but finite amount of $S$-mixing in each ring, and the macrospin approximation for these doesn't apply \cite{Candini10}. 

In conclusion, the present approach allows one to apply entanglement witnesses and spin-squeezing inequalities to the detection of entanglement between collective spins. The inequalities have been applied to a number of prototypical spin clusters, characterized by different geometries and coupling regimes. Here, entanglement between complementary subsystems is systematically detected in the ground state. The threshold temperature, above which thermal entanglement vanishes, is lower that the one obtained from the temperature dependence of the negativity, by a factor of the order of 2. Approximate expressions have been derived for the lower bounds in the inequality, that allow the derivation of all the relevant quantities from spin-pair correlation functions. The values of these expressions typically approaches from below that of the exact expressions in the parameter range of interest. We note that, while this relation cannot been taken from granted in general, the validity of the assumption underlying the approximation (i.e., small fluctuations of the partial spin sums) can always be verified experimentally. 

\section*{Acknowledgements}

This work has been financially support by the FIRB project RBFR12RPD1 of the Italian MIUR and by Fondazione Cariparma.


\begin{thebibliography}{34}
\bibitem[{not({\natexlab{b}})}]{notaA3}
\bibinfo{note}{The total spin ${\bf S}={\bf S}_A+{\bf S}_B$ can vanish only for
  $S_A=S_B$. The only probabilities $p(S_k^A,S_k^B)$ that can be finite for
  $\langle {\bf S} \rangle=0$ are thus correspond to $S_k^A=S_k^B$.}
\bibitem[{not({\natexlab{c}})}]{notaA2}
\bibinfo{note}{The relative difference between the averages $\overline{S_A}$
  and $\overline{S_B}$ and the corresponding theoretical maxima decreases with
  increasing $s$: for $s=1/2,1,3/2$, the value of
  $\overline{S_A}/S_M=\overline{S_B}/S_M$ in the ground state is $ 0.895,
  0.904, 0.935$, respectively. Also, the ground-state energy corresponding to
  the two-macrospin model approaches that of the spin ring.}
\bibitem[{not({\natexlab{d}})}]{notaA1}
\bibinfo{note}{The normalized value of the average spin length corresponding to
  each subsystems decreases with increasing $s$. In fact, for $s=1/2,1,3/2$,
  the value of $\overline{S_A}/S_M=\overline{S_B}/S_M$ in the ground state is $
  0.167, 0.155, 0.140$, respectively.}
\end{thebibliography}
\end{document}